\documentclass[pra,superscriptaddress,reprint]{revtex4-1}

\usepackage{amsmath,amssymb,graphicx,color,bm}
\usepackage{soul} %\st \hl

\begin{document}

\title{Observation of optical states below light cone with compound lattices}

\author{Jiajun Wang}
\affiliation{Department of Physics, Key Laboratory of Micro-and Nano-Photonic Structures (MOE), and State Key Laboratory of Surface Physics, Fudan University, Shanghai 200433, China}
\author{Ang Chen}
\email[Corresponding author:~]{achen12@fudan.edu.cn}
\affiliation{Department of Physics, Key Laboratory of Micro-and Nano-Photonic Structures (MOE), and State Key Laboratory of Surface Physics, Fudan University, Shanghai 200433, China}
\author{Maoxiong Zhao}
\affiliation{Department of Physics, Key Laboratory of Micro-and Nano-Photonic Structures (MOE), and State Key Laboratory of Surface Physics, Fudan University, Shanghai 200433, China}
\author{Wenzhe Liu}
\affiliation{Department of Physics, Key Laboratory of Micro-and Nano-Photonic Structures (MOE), and State Key Laboratory of Surface Physics, Fudan University, Shanghai 200433, China}
\author{Yiwen Zhang}
\affiliation{Department of Physics, Key Laboratory of Micro-and Nano-Photonic Structures (MOE), and State Key Laboratory of Surface Physics, Fudan University, Shanghai 200433, China}
\author{Xiaohan Liu}
\affiliation{Department of Physics, Key Laboratory of Micro-and Nano-Photonic Structures (MOE), and State Key Laboratory of Surface Physics, Fudan University, Shanghai 200433, China}
\affiliation{Collaborative Innovation Center of Advanced Microstructures, Fudan University, Shanghai 200433, China}
\author{Lei Shi}
\email[Corresponding author:~]{lshi@fudan.edu.cn}
\affiliation{Department of Physics, Key Laboratory of Micro-and Nano-Photonic Structures (MOE), and State Key Laboratory of Surface Physics, Fudan University, Shanghai 200433, China}
\affiliation{Collaborative Innovation Center of Advanced Microstructures, Fudan University, Shanghai 200433, China}
\author{Jian Zi}
\email[Corresponding author:~]{jzi@fudan.edu.cn}
\affiliation{Department of Physics, Key Laboratory of Micro-and Nano-Photonic Structures (MOE), and State Key Laboratory of Surface Physics, Fudan University, Shanghai 200433, China}
\affiliation{Collaborative Innovation Center of Advanced Microstructures, Fudan University, Shanghai 200433, China}

\begin{abstract}
 For optical systems, states inside the light cone could be detected by far-field measurement, while those below the light cone are not detectable by far-field measurement. A new method for far-field detection has been developed for observing states below the light cone with compound lattices. The basic mechanism involved is that periodic weak scattering leads to band folding, making the states out of the light cone to occur inside. By using polarization-resolved momentum-space imaging spectroscopy the band structures and iso-frequency contours of plasmonic lattices with different dimensions and symmetries are experimentally mapped out, in good agreement with the simulation.
\end{abstract}

\maketitle

%\section{Introduction}

Optical materials such as photonic crystals are attractive due to their fantastic ability of controlling and manipulating the flow of light~\cite{joannopoulos2011photonic}. Generally, band structures determine the optical response of such optical periodic structures, leading to novel phenomena of light propagation such as negative refraction~\cite{luo2002all,martinez2005negative}, super prism~\cite{kosaka1998superprism,amet2010experimental}, and self-collimation~\cite{yu2003bends, tang2006efficient}, etc. Besides, iso-frequency contours could offer an alternative way to describe light propagation from another perspective~\cite{shi2010direct}. Experimentally, near-field and far-field detection are usually carried out to get the band structures and iso-frequency contours. For near-field detection, amplitude and  phase information are imaged via a scanning probe tip. However, some unavoidable drawbacks limit its application. The preparation of special probe tips is difficult and complicated and the tip size hinders its application in frequency range of visible light. In addition, coherent light is required and only single frequency is allowed per scan. All these limitations lead to the low efficiency and restrict the use of near-field detection~\cite{dunn1999near, gerber2014phase}. Compared with near-field detection, far-field detection is much easier and more efficient. Even white light source could satisfy the measurement requirements, thus different frequencies in certain range could be detected simultaneously. Fourier transform of the lens results in the dispersion relation and iso-frequency contours directly~\cite{shi2010direct,regan2016direct,zhang2018observation,zhou2018observation}, however, it is limited by the coupling between modes inside optical structures and those in the free space, i.e. $k_x^2 + k_y^2 \le (\omega / c)^2$ is satisfied~\cite{fan2002analysis}, where $(k_{x}, k_{y})$ are components of the in-plane wave vector, $\omega$ the frequency and $c$ the speed of light in vacuum. The region satisfying $k_x^2 + k_y^2 \le (\omega / c)^2$ is called light cone.

The states below the light cone, i.e. $k_x^2 + k_y^2 > (\omega / c)^2$, have been an interesting topic for a very long time. For example, slow light at band edge below the light cone is researched for tunable time delays~\cite{povinelli2005slow}. Recently, researchers observed topological edge states below the light cone~\cite{wu2017direct}. However, states below the light cone could not be detected by far-field measurement. In this paper, a method of compound lattice is presented, in which the periodic weak scattering of compound lattice is used to transfer the states below the light cone to those inside the light cone, then band structures and iso-frequency contours below the light cone could be obtained by far-field detection.

\begin{figure*}[t]
\centering
\includegraphics[scale = 1.0]{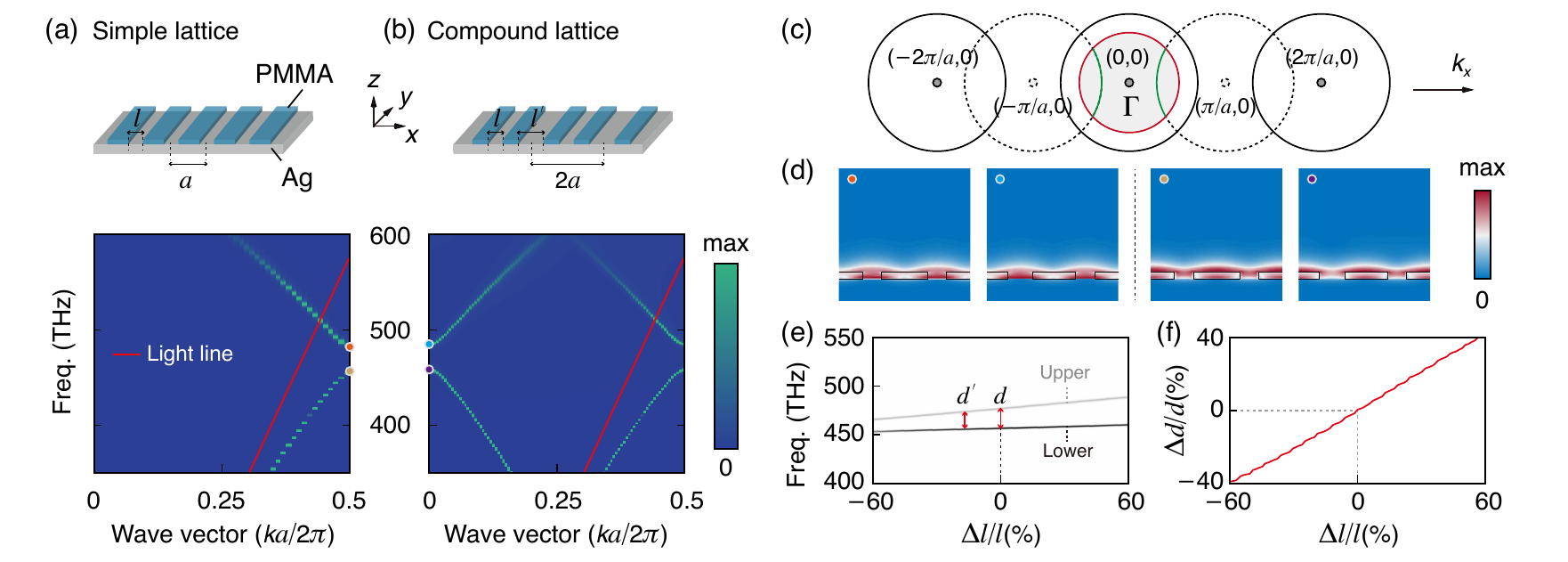}\\
\caption{\label{fig:1} Schematic views and corresponding band structures simulated along $x$-direction for (a) simple and (b) compound lattices of the one-dimensional PMMA grating on flat Ag substrate. The periods are $a$ and $2a$ for these two lattices respectively, with $a = 260$ nm. The width of air for (a) is $l = 70$ nm, while two widths of air for (b) are $l = 70$ nm and $l^{\prime} = 100$ nm. (c) Iso-frequency contours of the one-dimensional compound lattice: area inside the red circle corresponds to the cross section of light cone; solid black circles are the iso-frequency contours of the one-dimensional simple lattice, which are spaced by $2\pi /a$, the base vector of the reciprocal lattice. Dashed circles are iso-frequency contours folded by periodic weak scattering of the compound lattice. (d) Field distributions of the states represented by the four colored points(red,blue,yellow,purple) in (a) and (b). Nearly same fields for red (blue) and yellow(purple) points show the physical essence of the periodic weak scattering. (e) Positions of the two bands at the boundary of BZ zone. $\Delta l$ is the variation of the width of air changed. The corresponding relative change of band gap $\Delta d/ d =(d^{\prime} - d)/ d$ is shown in (f).}
\end{figure*}

The structure studied is a flat Ag film covered with a patterned thin film of Poly(methyl methacrylate)(PMMA), in which the surface plasmon polaritons (SPPs) show well-defined band structures, and modes above the light cone can couple into the free space and radiate~\cite{han2006transmission}. With elaborate designs of the compound lattice, the bands below the light cone could be folded to be measurable by far-field detection. The outline of the paper is as follows. First, the simulation and experimental results of one-dimensional (1D) plasmonic lattice are presented. Experimental results are compared with finite-difference time-domain (FDTD) simulation. Next, the experimentally mapped band structures with iso-frequency contours of two-dimensional (2D) lattices are shown along different directions. Finally the conclusion is presented.

%\section{One-dimensional compound lattices~\label{sec:one-dimensional}}

\begin{figure}[t]
\centering
\includegraphics[scale = 1.0]{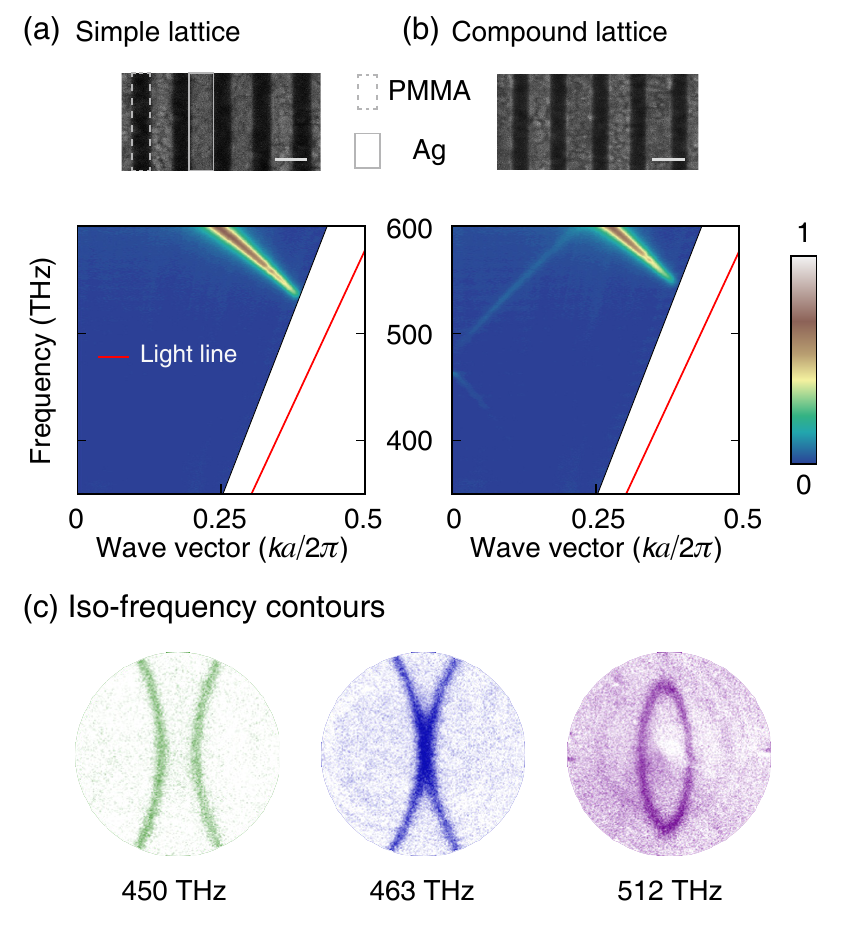}\\
\caption{\label{fig:2} SEM images and corresponding band structures (extinction spectra) of simple lattice (a) and its compound counterpart (b), obtained experimentally. The scale bar is 500 nm. White regions correspond to the states out of the numerical aperture of the microscope. To compare with the simulation results, the light line is also shown. (c) Three typical iso-frequency contours with 450, 463 and 512 THz for the compound lattice show the evolution of iso-frequency contours when the frequency increases. The green one corresponds to the case shown in Fig. 1(c).}
\end{figure}

The basic principle of the present method is explained with an one-dimensional plasmonic lattice shown in Fig. 1. The band structures of both the simple lattice and its compound-lattice counterpart are simulated by using FDTD method. The sample geometry is: for simple lattice the width of each PMMA stripe is 190 nm  while the length is much longer than its width and could be considered as infinity, and the width of air channel $l$, i.e. separation between two adjacent stripes is 70 nm, making period of the so-called 1D simple lattice equal to $a=260$ nm, as shown in Fig. 1(a). While for the compound-lattice counterpart, every two primitive cells of the simple lattice is set as an unit where two kinds of air channels are included. The width $l$ of one kind of air channel is kept unchanged, while the width of the other kind of air channel is tuned from $l=70$ nm to $l^{\prime} = 100$ nm. Therefore the period of the compound lattice is doubled as $2a=520$ nm, as shown in Fig. 1(b). Due to the periodic weak scattering of the compound lattice, the first-order band of the simple lattice below the light cone is folded inside the light cone, as shown in Fig. 1(b). A schematic view of iso-frequency contours is shown in Fig. 1(c) for the compound lattice, providing the information about the principle of the band folding caused by periodic weak scattering. Solid black circles are the iso-frequency contours of the one-dimensional simple lattice, which is spaced by $2\pi /a$. It can be seen that additional dashed contours occur whose centers are $\pi /a$ from $\Gamma$ point, for reciprocal lattice vector of periodic weak scattering of the compound lattice is $\pi /a$. Green curves could be detected by far-field since they have fallen into the light cone, whose boundary is denoted by the red circle in Fig. 1(c). To compare the folded bands with those of the simple lattice below the light cone, the field distributions of the upper and lower bands are calculated at $k = \pi /a$ for simple lattice and at $k = 0$ for compound lattice, as shown in Fig. 1(d). They are corresponding to the four colored points(red,blue,yellow,purple) in Fig. 1(a) and (b). Nearly the same distribution for both upper and lower bands shows the validity of the present method. Moreover, the influence of periodic weak scattering is investigated by comparing the width change of the band gap between the upper and lower bands. Changing the width of air channel from $l$ to $l^{\prime}$, the relation between the width of band gap and the relative width change of air channel $\Delta l / l = (l^{\prime} - l) / l$ is also shown in Fig. 1(e), where $d$ is the band gap of the simple lattice. The width of band gap increases with $l^{\prime}$. Fig. 1(f) shows the relation between the relative change of the band gap $\Delta d /d =(d^{\prime} - d) / d$ and the relative change of the air channel width $\Delta l / l$. Another important information could also be found, i.e. the lower band changes less than the upper band as shown in Fig. 1(e).

\begin{figure}[t]
\centering
\includegraphics[scale = 1.0]{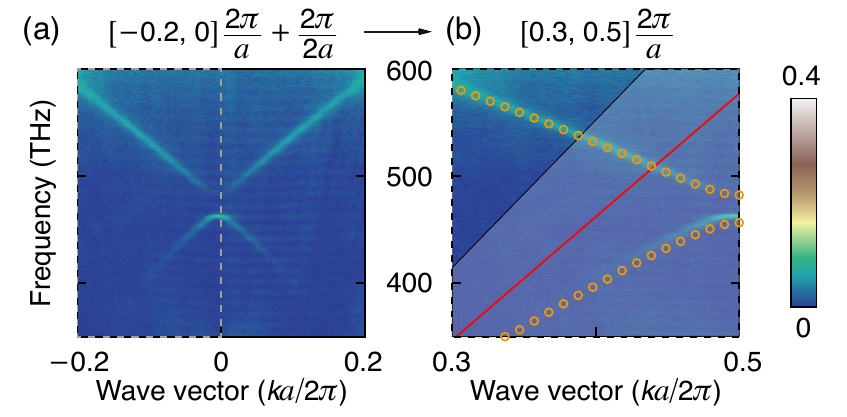}\\
\caption{\label{fig:3} (a) Zoomed band structures of compound lattice shown in Fig. 2(b). Band structures with negative wave vectors are also shown. A reciprocal lattice vector along $k_{x}$ direction of the compound lattice is $2 \pi /2a = \pi/a$. (b) The obtained band structure of the 1D simple lattice, with the simulation results denoted by orange circles for comparisons.}
\end{figure}

To experimentally characterize the band structures and iso-frequency contours, plasmonic lattices are fabricated by using electron-beam lithography. The Ag substrate is thick enough to avoid transmission of the incident light such that the extinction spectra $E_{\text{xt}} = 1 - R$ could be used to characterize the bands and contours. By using the homemade polarization-resolved momentum-space imaging spectroscopy~\cite{zhang2018observation}, both band structures and iso-frequency contours could be obtained experimentally, as shown in Fig. 2. The SEM images together with their corresponding band structures are shown in Fig. 2(a) and (b) for simple lattice and its compound counterpart, respectively. The red line is the light line, satisfying $k_x^2 + k_y^2 = (\omega / c)^2$. The white regions above the red line represents the states which could not be collected by the microscope due to its limited numerical aperture. The states interested are just inside the white regions. To check if the band folding were caused by periodic weak scattering, the iso-frequency contours of the compound lattice for several different frequencies are further measured. With the frequency increasing, two arcs (at 450 THz) corresponding to Fig. 1(c) get touched (at 463 THz) and finally form a closed curve (at 512 THz), as shown in Fig.2(c).

\begin{figure*}[t]
\centering
\includegraphics[scale = 1.0]{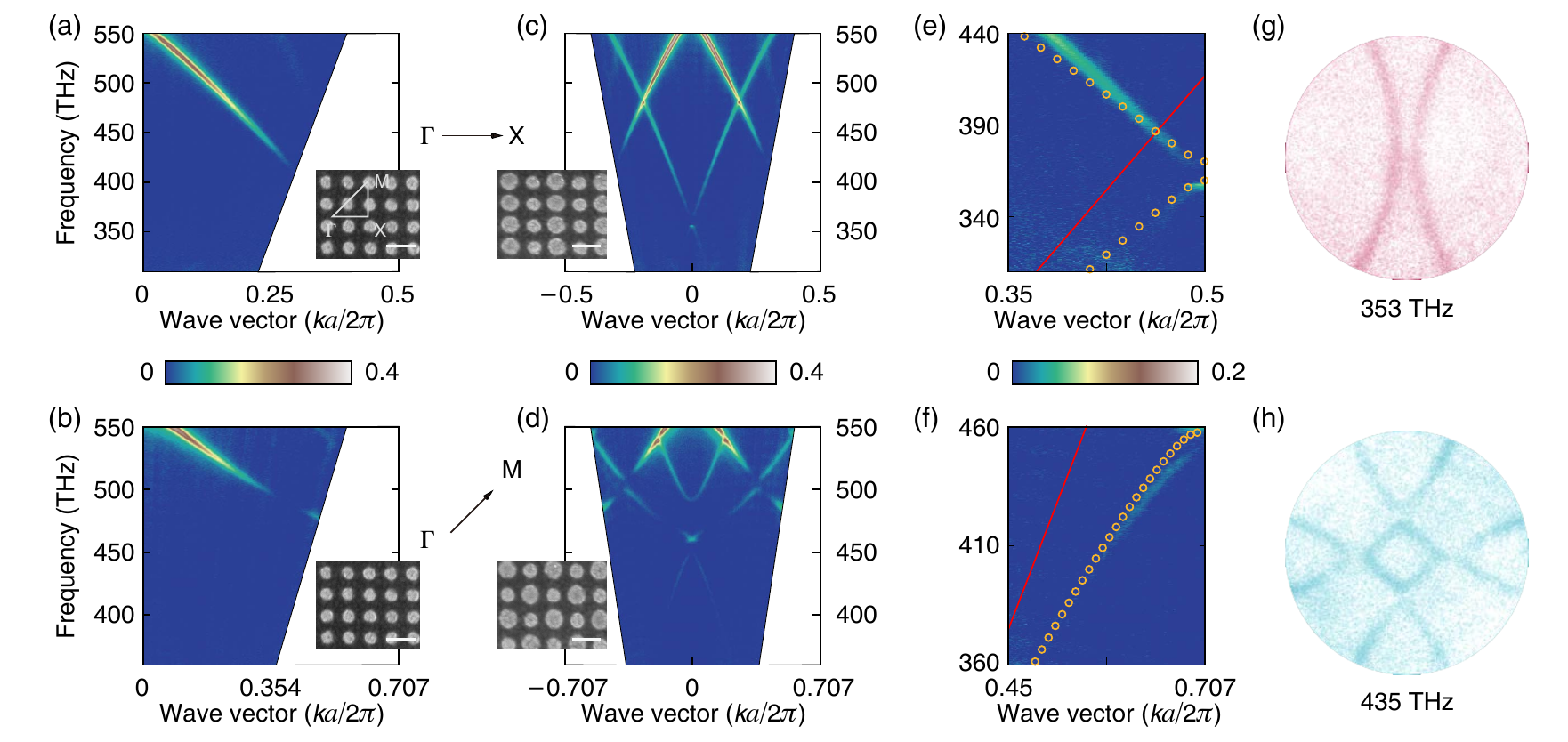}\\
\caption{\label{fig:4} Experimental band structures along $\Gamma$-X (a) and $\Gamma$-M (b) of 2D square lattice. The lattice constant is 360 nm. The radius of air circle is 90 nm. (c) and (d) show the measured band structures for two designed compound lattices for the two symmetry directions. The radius of the bigger air holes is 120 nm, while that of the smaller ones is fixed to 90 nm. The scale bar is 500 nm. (e) and (f) are the obtained bands of simple 2D square lattice, with the simulated results denoted by orange circles. (g) and (h) give the corresponding iso-frequency contours at the frequencies 353 and 435 THz, respectively.}
\end{figure*}

The band structures of the simple lattice below the light cone could be reconstructed by the measured band structures of its compound-lattice counterpart. Fig. 3(a) shows the zoomed bands (including negative wave vectors) of the compound lattice shown in Fig. 2(b). The reciprocal vector for the compound lattice along the $k_{x}$ direction becomes $\pi/a$ rather than $2\pi/a$ for the simple lattice, thus the bands, for instance, within $(-0.2, 0)2 \pi/a$ could be translated to those within $(0.3, 0.5)2 \pi/a$ by one reciprocal vector $\pi/a$ transition. The obtained band structure of the 1D simple lattice is shown in Fig. 3(b), in good agreement with the simulation of simple lattice by FDTD method, denoted by orange circles in Fig. 3(b).

%\section{Two-dimensional compound lattices~\label{sec:two-dimensional}}

Next the case of 2D plasmonic lattices is considered. The bands are usually mapped along directions of high-symmetry. Besides the translational symmetry, the specific symmetries for higher-dimensional lattices always have more significance. For simplicity, a 2D square lattice is taken as an example to show the present method. Fig. 4(a) and (b) show the two experimental band structures along $\Gamma$-X and $\Gamma$-M of the simple lattice, together with the SEM images, where X point and M point are at ($\pi/a,0$) and ($\pi/a,\pi/a$) in reciprocal space respectively. This simple lattice is a 2D square lattice consisting of an array of round air holes. The period of the simple lattice is 360 nm and the radius of each air hole is 90 nm. The symmetry of point X and M is different that X is of the $C_{2v}$-symmetry while M the $C_{4v}$~\cite{joannopoulos2011photonic}. Thus, two different designs of compound lattices as the counterparts of the square simple lattice are required to observe the bands along these two directions. First for $\Gamma$-X direction, the compound lattice should keep the $C_{2v}$-symmetry. The SEM image shown in the inset of Fig. 4(c) ensures that the bands can be folded along $\Gamma$-X direction where the small air holes keep the same radius 90 nm as those of the simple lattice while the radius of the bigger ones is changed to 120 nm. As for $\Gamma$-M direction, the compound lattice should be designed to keep $C_{4v}$-symmetry, which is shown in Fig. 4(d) where the small air holes keep the same radius as the simple lattice while the radius of the bigger ones is changed to 120 nm. The obtained band structures of the simple 2D square lattice for those two directions are in good agreement with the simulation of the simple lattice. Typical iso-frequency contours for the two types of structures are shown in Fig. 4(g) and (h). The symmetry information along different directions could also be observed from the iso-frequency contours.

By different designs for different high-symmetry directions, the band structures of the simple 2D square lattice below the light cone could be obtained by far-field detection. Therefore the method presented in this paper could be generally applied to many other symmetric structures such as triangular or honeycomb lattices.

%\section{Conclusion~\label{sec:conclusion}}

In conclusion, the feasibility and validity of the method with compound lattice to observe the optical states below the light cone has been analyzed. The periodic weak scattering leads to band folding, then band structures and iso-frequency contours could be measured with far-field detection system. States below the light cone are obtained experimentally. This method provides an efficient and practical option to study the state below the light cone.

%\begin{acknowledgments}

The work was supported by 973 Program and China National Key Basic Research Program (2015CB659400, 2016YFA0301100 and 2016YFA0302000) and National Science Foundation of China (11774063 and 11727811). The research of L. S. was further supported by Science and Technology Commission of Shanghai Municipality (17ZR1442300, 17142200100) and the
Recruitment Program of Global Youth Experts (1000 plans).

%\end{acknowledgments}

\bibliography{observing_states_below_line_ref}

\end{document}